# Discovery of probably Tunguska meteorites at the bottom of Khushmo river's shoal

Andrei E. Zlobin

*Vernadsky State Geological Museum, Russian Academy of Sciences*
*Mokhovaya 11/11, 125009, Moscow, Russian Federation*
*e-mail: z-tunguska@yandex.ru*

*Abstract.* The author describes some stones which he found at the bottom of Khushmo River's shoal during 1988 expedition into the region of the Tunguska impact (1908). Photos of stones are presented. Three stones have traces of melting and the author consider these stones as probable Tunguska meteorites. Some arguments are presented to confirm author's opinion. Results of investigation of prospect holes in peat-bogs are briefly described too. New data concerning heat impulse of the Tunguska impact are obtained. There is the assumption that some meteorites which are formed during comet impact looks like stony or glass-like thin plates with traces of melting.

*Keywords*: Tunguska, Tunguska meteorite, Tunguska bolide, Tunguska fire-ball, Tunguska explosion, Tunguska comet, Tunguska catastrophe, Tunguska event, Tunguska impact, comet, meteorite, stony meteorite, regmaglypts, shatter cones, heat transfer, heat impulse, explosion, supersonic, hypersonic, gas dynamics, gas flow, vortex, hollow, depression, stagnation point, jet, temperature stresses, cryogenics

## 1. Introduction

In accordance to UNESCO information, 2013 is declared as the year of 150th anniversary of the birth of Vladimir Ivanovich Vernadsky (March 12). Academician V.I.Vernadsky is well known as famous scientist and main supporter of L.I.Kulik's expeditions into the region of the Tunguska impact. Vernadsky and Kulik were considering very important to find new meteorites and to deliver these samples into the collection of Russian Academy of Sciences [Vernadsky, 1921]. Academician N.V.Vasilyev made a lot to organize investigation of Tunguska impact with high academic level too [Vasilyev, 2004]. Today seems necessary to remind about some historical facts related to investigation of the Tunguska event, and to describe some samples of substance discovered by the author at the bottom of Khushmo River's shoal in 1988.

Scientists try to find fragments or sediments of the Tunguska space impactor approximately during 100 years. It is known that a piece of glass substance with bubbles (0.5 kg) was already found near the Suslov's depression during expedition of L.A.Kulik. Originally Kulik considered this melted glass sample as related to the Tunguska event [Kulik, 1939]. Later, that piece of glass was lost and Kulik's opinion was not taken into consideration during long time. However, till now a part of scientists don't exclude possibility, that considerable fragments of the Tunguska meteor body may be found [Vasilyev, 2004].

## 2. Traces of thermal influence

It is known that powerful thermal influence took place during the Tunguska impact and branches of trees were heated and burnt. The author carried out special experimental investigation for determination of thermal properties of tree's rind and blast heat impulse. Obtained thermal properties were used for calculations of temperature distribution in cross section of branches. Heat impulse 13 - 30 $J/cm^2$ was determined during analysis with 2-D finite element method [Zlobin, 2007]. Such level of heat impulse was able to heat branches of trees, but was not able to make melting of stones on the ground. Heat impulse during the Tunguska explosion was also estimated as 26 - 34 $J/cm^2$ without effect of complete water evaporation from tree's needles [Vaganov et al., 2004].

In 1988 the author of this paper participated expedition into the region of the Tunguska impact. During this expedition he investigated influence of heat impulse and thermal damages of vegetation's sediments caused by the Tunguska impact. After quasi three-dimensional modeling [Zlobin, 2007] it became possible to compare view of real thermal influence in the site and results of calculated heat impulse. The author made more than ten prospect-holes in the peat-bogs including the prospect-hole near the Suslov's depression [Fig. 1]. Other places were: Southern peat-bog, peat-bog near Laboratory camp (not far from Cheko Lake), several peat-bogs along Western Section, Bublik peat-bog and another places around central region of the Tunguska impact. All layers of peat-bogs were closely inspected, including layers in the depth of permafrost which dated as 1908. The layer of fire of 1908 was detected accurately. However, any presence of considerably pieces of glass-like substance was not discovered in prospect-holes. As the result, the author decided to search melted substance of the Tunguska space body on the bottom of shoal of river, where natural cleaning of stones is produced by clean water.

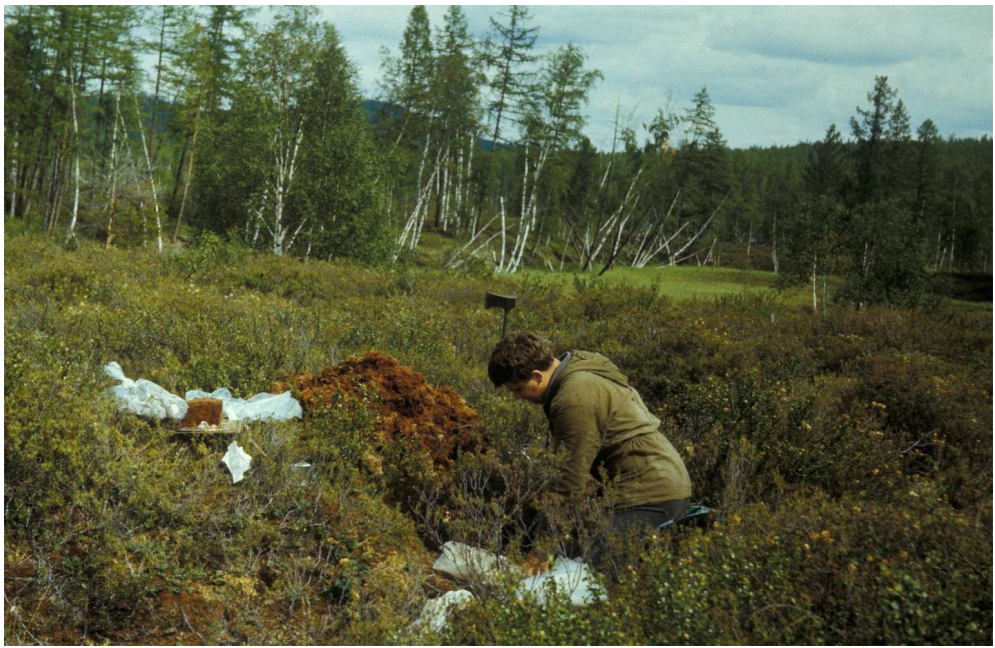

Fig. 1. The author makes prospect-hole in the peat-bog not far from Suslov's depression (central region of the Tunguska impact, July 1988).

## 3. Stones with traces of melting

During the expedition of 1988, in July 24 the author arrived at Pristan camp near the coast of the Khushmo River. He was there from July 24 to July 26. Before returning to Kulik's Zaimka main camp, the author investigated the shoal of the Khushmo River near Pristan with the purpose to find stones which looks like meteorites. Also some stones were collected which seems as aesthetic. Stones were good visible at the bottom of the shoal and the author found interesting samples. All collection consisted of more than 100 stones and the author delivered these samples to Moscow by airplane. Gross weight of all stones was approximately 1.5 kg.

After the expedition the author focused his efforts on experimental investigation of thermal processes and mathematical modeling of the Tunguska impact [Zlobin, 2007]. It was shown during quasi three-dimensional modeling that small heavy fragments of the Tunguska impactor probably were dissipated in several points with concrete coordinates, including region of Pristan camp. The temperature of final flare was calculated too (1700 K) which indicated possibility of melting of some stony-like substance into the volume of fire-ball. In 2008 the author sorted his collection of stones from the Khushmo River's shoal and selected three stones with traces of melting, which were described and officially registered.

Let us give more detailed description of mentioned three stones. General view of stones is presented at Fig. 2, Fig. 3 and Fig. 4. The author gave names to stones for more convenient description of their features - "dental crown", "whale" and "boat". From the view of maximal surface area all three stones has the form approximately like parallelogram. Measured size and weight of stones are presented in Table 1. The "whale" stone considerably more massive than other two stones and its base looks approximately flat. There are good visible traces of melting on the surface of all stones. Moreover, stones has surface structures which looks like regmaglypts and the "boat" stone has deep cavern. Colors of stones are: rusty-brown and yellow ("dental crown"), dark brown ("whale") and rusty-brown ("boat"). Traces of thin bubble-like structures are visible on concave surface of "dental crown" stone.

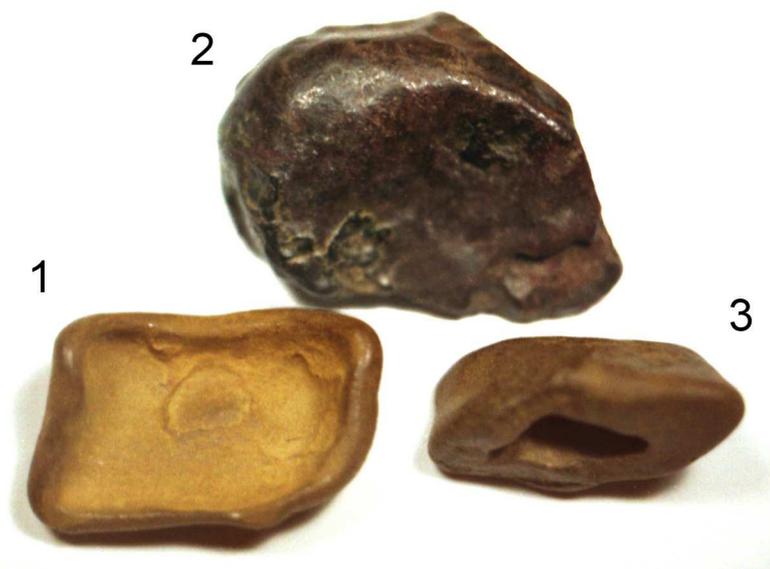

Fig. 2. General view of stones:  1 - Dental crown,  2 - Whale,  3 - Boat

Table 1

|  | Dental crown | Whale | Boat |
|---|---|---|---|
| Maximal diagonal (mm) | 25 | 29 | 21 |
| Weight (g) | 1.6 | 10.4 | 2.3 |

There is good visible imprint of impact of another body on concave surface at melted edge of "dental crone" sample (interaction between plenty of fragments). Also it is necessary to note very interesting structure on convex side of "dental crown" sample (Fig. 3). This structure looks like not deep spherical depression with tracks of solidification of couple of liquid vortexes at the bottom of the depression. Such structure seems possible only in case of heating of convex side of the sample by hot gas flow with tangential component of velocity at the region of stagnation point. The same effects of vortex flows at surface with hollow-type relief are well known phenomena [Leontiev et al., 2002]. The influence of tangential velocity component of hot gas flow seems visible in general view of each of three stones. This confirms author's assumption that these stones obtain traces of melting during powerful explosive expansion of fire-ball's volume.

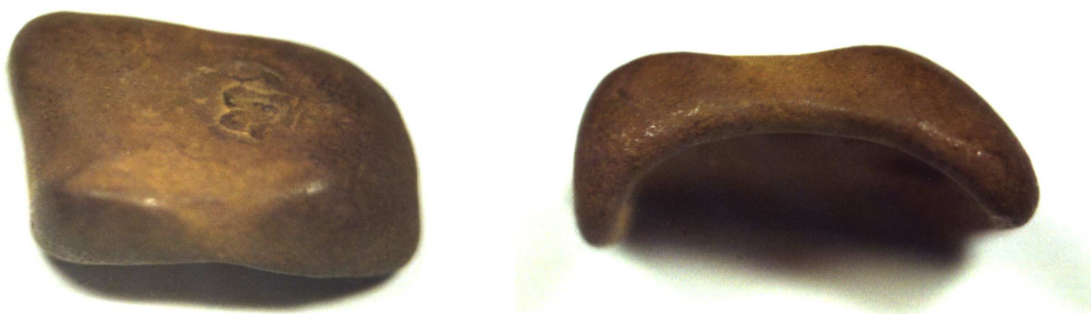

Fig. 3. "Dental crown" stone with spherical depression on convex side and tracks of solidification of couple of liquid vortexes

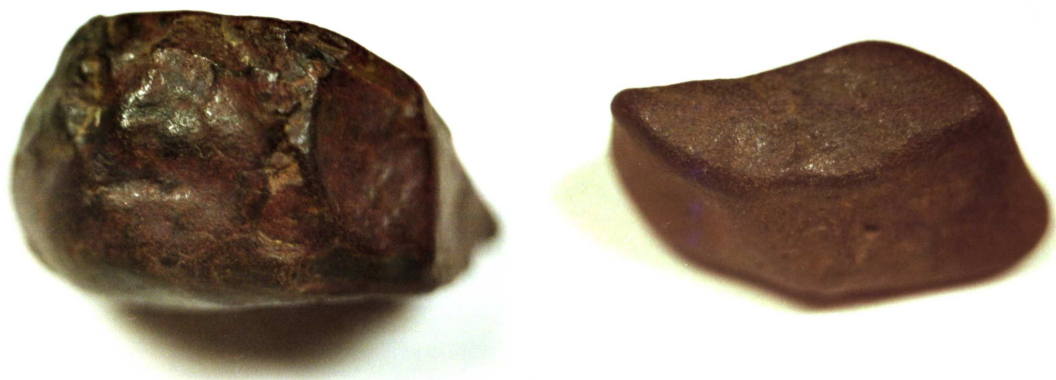

Fig. 4. Damages on the surface of "whale" stone and melted collar flange on the surface of "boat" stone

## 4. Heat for melting

The form of "dental crown" stone (thin plate) is ideal for thermal measurements. The author investigated process of heating of this thin plate to determine conditions of plastic deformation and melting for "dental crown" sample. Before calculations the assumption was done that the sample consists of quartz-like substance ($SiO_2$). It was determined during calculations how much heat is required for softening and external melting of quartz-like sample. The equation of thermal balance was used the same as for micro-meteorites [Whipple, 1950, 1951], when thickness of meteorite considerably less than 1 cm. Calculations was carried out for several cases of initial and boundary conditions with next generalization of results. During mathematical modeling the author determined necessary heat impulse between 280 and 420 $J/cm^2$. This is considerably higher than values 13 - 30 $J/cm^2$ which were determined earlier for the level on the ground [Zlobin, 2007]. That is why the author considers that stones with traces of melting could be heated not on the ground, but directly in the volume of the Tunguska fire-ball, during motion through the atmosphere in considerably altitude (Fig. 5). Moreover, heat impulse between 280 and 420 $J/cm^2$ is in good correspondence to conditions of astroballistic heat transfer [Kutateladze, 1970]. Certainly it will be possible to determine more accurate value of heat impulse, necessary for melting of stones, after chemical analysis of the substance.

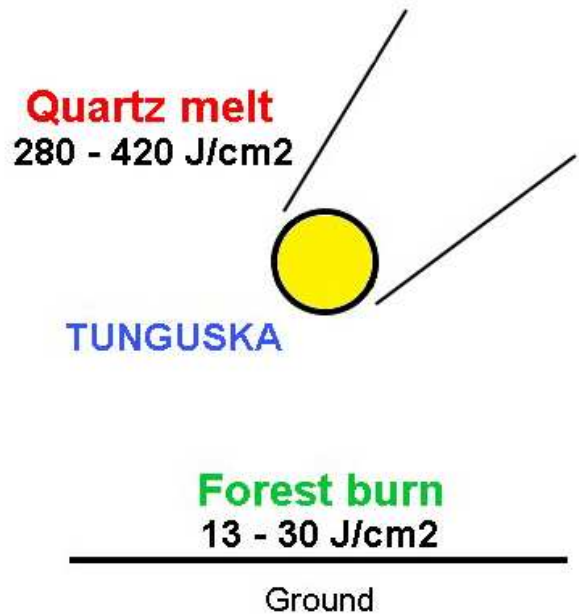

Fig. 5. Comparison between heat impulse in the fire-ball on considerable altitude and heat impulse on the ground

Let us notice that upper level of total explosion heat impulse 300 $J/cm^2$ was also estimated from position of complete water evaporation from tree's needle [Vaganov et al., 2004]. If to say concerning trees damages, Vaganov et al. concluded that unlikely heat impulse exceeded 300 $J/cm^2$ (no signs of tree's crown fire) and the minimum heat impulse was estimated by these authors as 25 $J/cm^2$.

## 5. Other stones of collection

Most of another stones from Khushmo collection of the author are presented at Fig. 6. Stones have different size and color and some of them confirm the fact of powerful impact event too. The author found "shatter cones" (Fig. 7) not far from melted stones. The size of "shatter cones" is between 21 and 16 mm.

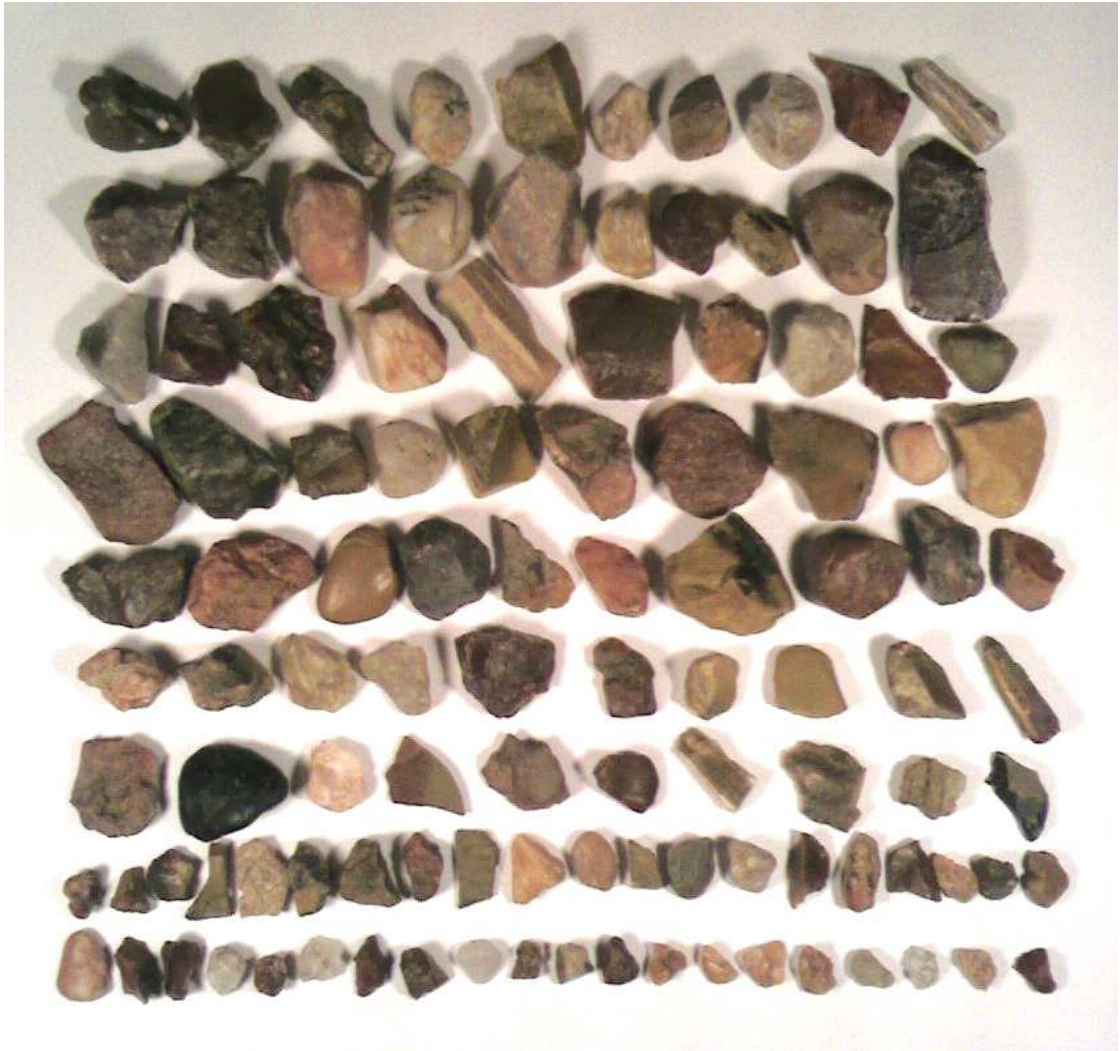

Fig. 6. Other stones of author's collection from Khushmo River's shoal

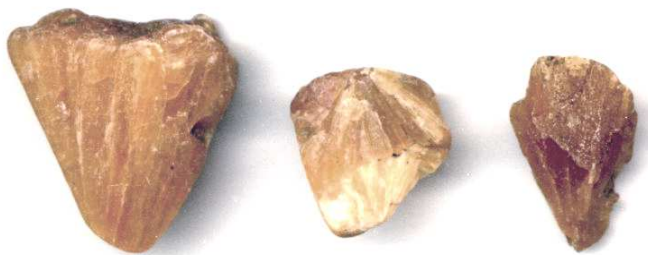

Fig. 7. Shatter cones

## 6. Conclusion

There are a lot of arguments which confirm the discovery of Tunguska meteorites:

- three melted stones with the size of ~20 - 30 mm were discovered near the Pristan region, which was marked during mathematical modeling as probably area with meteorites fall (relative coordinates X=0.321, Y=0.272) [Zlobin, 2007];
- stones has surface structures which looks like regmaglypts;
- it is not excluded that the form of parallelogram of melted stones means common origin of samples from position of similar initial structure of crystals. Therefore, melted stones is possible to examine as fragments of initially common stony body;
- traces of melting on surfaces of stones demonstrates influence of intensive heat impulse which is possible in the volume of giant bolide (280 - 420 $J/cm^2$);
- "dental crown" stone looks like a piece of melted glass. It is known a piece of glass was found near the Suslov's depression during expedition of L.A.Kulik too [Kulik, 1939]. Both these samples are similar if to note presence of traces of bubbles;
- tracks of solidification of couple of liquid vortexes on the surface of "dental crown" stone confirm high speed motion of this body in the volume of high temperature gas with considerably tangential component of velocity (intensive expansion of volume during explosion-like process). The same influence of tangential component of velocity seems visible on forms of "whale" and "boat" stones;
- there is good visible imprint of impact of another body on concave surface at melted edge of "dental crone" sample. This imprint was able to appear in case of interaction between plenty of fragments of the Tunguska body during explosive-like destruction in the atmosphere;
- rusty-brown color of stones possible signalize concerning presence of some quantity of iron [Fe], which is known as component of meteorites too;
- the view of all three stones is considerably differ from another stones collected at the same place;
- the presence of "shatter cones" is known as attribute of impact events [Melosh, 1989].

Certainly, strict confirmation of discovered melted stones as Tunguska meteorites is possible only after attentive chemical analysis of substance. Possible discovering of stony meteorites not excludes a comet as main ice mass of the Tunguska impact. After quasi three-dimensional modeling the author has already demonstrated, that average density of the Tunguska space body was 0.6 $g/cm^3$ [Zlobin, 2007]. This density of comet nucleus is in good correspondence to obtained density of Halley comet [Sagdeev et al., 1988]. Moreover, mathematical model of ice comet considerably better explain intensive destruction of small stony bodies in the atmosphere. Compound structure of the Tunguska comet's nucleus seems possible. If small stony bodies initially are hidden in comet's ice, these bodies are in condition of ultra low temperature [Zlobin, 1995]. During intensive destruction of comet's nucleus in the volume of fire-ball, stony bodies are being exposed to influence of ultra high and ultra low temperatures simultaneously. High level of temperature stresses are being realized near surface layer of stony bodies. Separation of small size thin stony plates due to temperature stresses may accompany destruction of stones. As a result, thin stony plates are able be dissipated completely due to mechanism of melting. The author pay attention on "dental crown" thin melted stony plate, which seems as excellent confirmation of cometary origin of the Tunguska impact.

## Acknowledgments

I very thankful to administration and my colleagues in Vernadsky State Geological Museum (RAS) on possibility to work with information concerning L.A.Kulik's activity and to analyze meteorites of the Museum.